\documentclass[notitlepage,twocolumn,superscriptaddress,amsmath,amssymb,aps,prl]{revtex4-2}
\usepackage{amsmath,amssymb,graphicx,mathtools,dsfont,ulem,bbold}
\usepackage[dvipsnames]{xcolor}
\usepackage[T1]{fontenc}
\usepackage{hyperref}
\usepackage{soul}
\usepackage{bibunits}
\hypersetup{citecolor=red,colorlinks=true,urlcolor=blue}

\newcommand{\ket}[1]    {|#1 \rangle}
\newcommand{\ketbra}[2]{|#1\rangle\!\langle#2|}

\newcommand{\tr}[1]    {{\rm Tr}\left[ #1 \right]}
\newcommand{\av}[1]    {\langle #1 \rangle}

\newcommand{\modsq}[1]    {\left| #1 \right|^2}
\newcommand{\modsqs}[1]    {\big| #1 \big|^2}

\newcommand{\En}{\mathcal E_N}

\newcommand{\ICFO}{ICFO - Institut de Ciencies Fotoniques, The Barcelona Institute of Science and Technology, 08860 Castelldefels, Barcelona, Spain}
\newcommand{\ICREA}{ICREA, Pg. Lluís Companys 23, 08010 Barcelona, Spain}
\newcommand{\IOPPAS}{Institute of Physics PAS, Aleja Lotnikow 32/46, 02-668 Warszawa, Poland}
\newcommand{\UW}{Faculty of Physics, University of Warsaw, ul. Pasteura 5, PL-02-093 Warsaw, Poland}
\newcommand{\TU}{Institute of Physics, Faculty of Physics, Astronomy and Informatics, Nicolaus Copernicus University
in Toru\'n, Grudzi\c{a}dzka 5, 87-100 Toru\'n, Poland }

\begin{document}

\title{Generation of scalable many-body Bell correlations  in spin chains \\ with short-range two-body interactions}

\author{Marcin P\l{}odzie\'n}\affiliation{\ICFO}\email{marcin.plodzien@icfo.eu}
\author{Tomasz Wasak}\affiliation{\TU}
\author{Emilia Witkowska}\affiliation{\IOPPAS}
\author{Maciej Lewenstein}\affiliation{\ICFO}\affiliation{\ICREA}
\author{Jan Chwede\'nczuk}\affiliation{\UW}\email{jan.chwedenczuk@fuw.edu.pl}

\begin{abstract}
 Dynamical generation of strong and scalable quantum resources, like many-body entanglement and Bell correlations, in spin-$1/2$ chains is possible with all-to-all interactions, either for constant interaction {strength} realizing one-axis twisting {protocol} or for power-law decaying {potentials}. We show, however, that such quantum resources can also be dynamically generated with a finite range of interactions.  We identify a necessary critical range and indicate a critical time when scalable quantum correlations appear. Finally, we show that the certification of generated states is accessible in the modern quantum simulator platforms. 
\end{abstract}

\maketitle

{\it Introduction}---
The {potential for} future quantum technologies {is} fuelled by quantum resources which are the quantum coherence and the triad of many-body non-classical correlations: entanglement~\cite{horodecki2009entanglement}, Einstein-Podolsky-Rosen (EPR) steering~\cite{uola2020steering} and the Bell nonlocality~\cite{brunner2014bell}. These are crucial for quantum computing, cryptography, communication and metrology---the key pillars of the second quantum revolution \cite{Acin_2018,Eisert2020,Fraxanet2022}. 
As such, the main goal of quantum technologies in the next years is 
generation, characterization, storage, and certification of many-body quantum states \cite{Frerot_2023}.

The archetypal model allowing generation of scalable quantum resources is a spin-$1/2$ chain undergoing one-axis twisting (OAT) \cite{kitagawa1993squeezed,wineland1994squeezed}. In the OAT protocol, the dynamics is governed by a non-linear Hamiltonian with all-to-all infinite range spin couplings, like $\hat{S}^2_z$, where $\hat{S}_z$ is a collective spin operator along $z$-axis, while initially spins are polarized along a direction perpendicular to the $z$-axis.
The OAT protocol dynamically generates quantum states such as spin-squeezed states useful for high-precision metrology limit~\cite{RevModPhys.90.035005,PhysRevLett.105.053601,PhysRevA.50.67,MllerRigat2023}, many-body entangled and the many-body Bell correlated states~\cite{Tura1256,schmied2016bell,Aloy2019,Baccari2019,Tura2019,PRXQuantum.2.030329,zukowski2002bell,cavalcanti2007bell,he2011entanglement,cavalcanti2011unified,spiny.milosz,PhysRevLett.126.210506,10.21468/SciPostPhysCore.5.2.025,PRXQuantum.2.030329, PhysRevA.46.R6797,PhysRevA.47.5138,PhysRevA.92.043622,PhysRevLett.100.210401,Li2009,PhysRevA.96.013823,Kajtoch-sc-2018, Schulte2020,PhysRevLett.126.160402, PhysRevA.105.022625}. 

The OAT dynamics can be realized with a variety of quantum simulating platforms, like ultracold systems utilizing atom-atom collisions~\cite{Treutlein2010,Oberthaler2010,Chapman2012,PhysRevLett.125.033401}, atom-light interactions~\cite{PhysRevLett.104.073602,PhysRevLett.105.080403}, in Rydberg atoms~\cite{doi:10.1126/science.aax9743, bornet2023scalable, Eckner2023}, array of trapped ions~\cite{Bohnet1297,franke2023quantumenhanced}, and superconducting qubits \cite{PhysRevLett.119.180511, doi:10.1126/science.aay0600, doi:10.1126/sciadv.aba4935}.
 The recent theoretical proposals for the OAT simulation with ultracold atoms in optical lattices effectively simulate Hubbard and Heisenberg models~\cite{Kajtoch2018, PhysRevResearch.1.033075, Plodzien2020, Plodzien2022, Plodzie2023generation, PhysRevResearch.3.013178,Hernandez2022,Dziurawiec2023,yanes2023spin}. All the above mentioned proposals are based on effectively induced all-to-all infinite-range couplings, either with constant amplitude as required by OAT, or with power-law decaying amplitude like dipolar or van der Waals interactions.

Here, we show that with short-range interactions, scalable many-body entanglement and Bell correlations can be generated with spin chains with Ising-type Hamiltonian.
In contrast to the long-range interactions and all-to-all couplings, in our system, the number of connected spins by the interaction, i.e., the number of two-qubit entangling operations, increases linearly with the system size for a fixed interaction range, which is an advantage compared to the former systems, which require quadratic scaling of the number of operations for the protocol.

\begin{figure}[t!]
    \centering
\includegraphics[width=\linewidth]{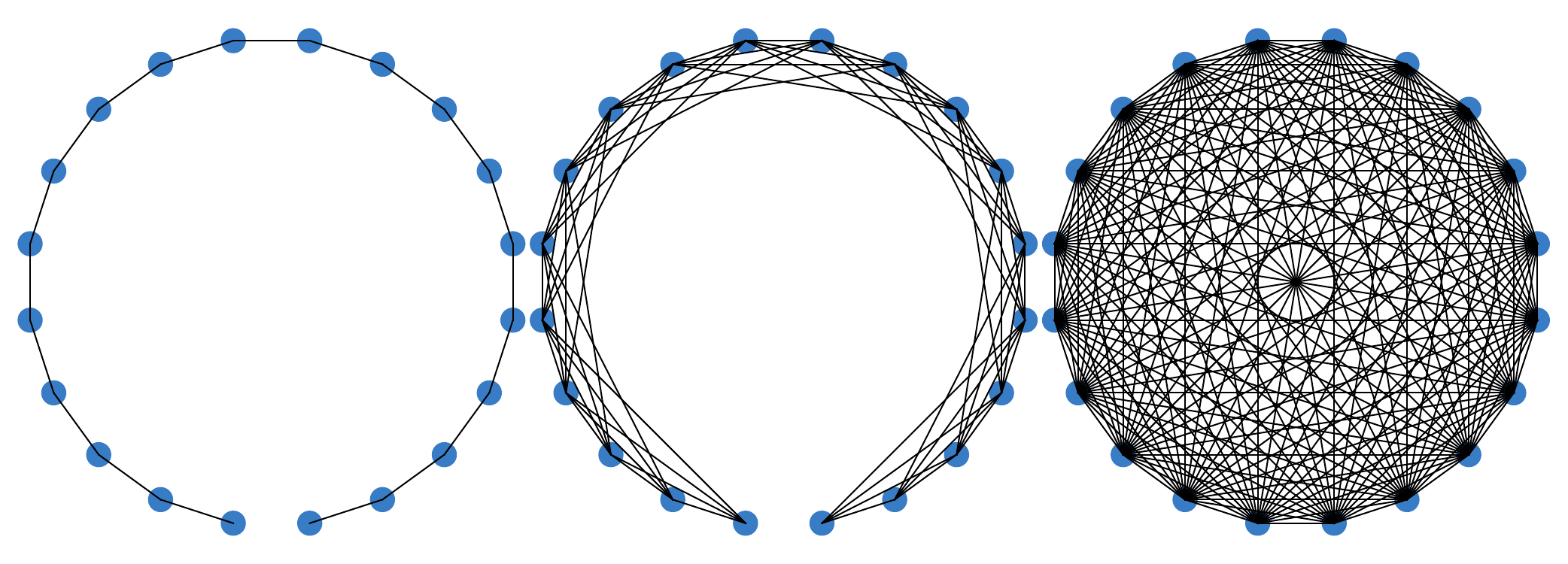}
    \caption{Visualization of the spin-connectivity in a chain of $N$ spins for different interaction ranges $r$. The three panels show nearest-neighbors interaction with the range $r=1$ (left), 
      finite-range interaction with $r=4$ (middle), and the all-to-all type of coupling with $r = N-1$, present in one-axis twisting-like protocols (right).}
    \label{fig.scheme}
\end{figure}
We find a critical range allowing many-body entanglement generation,  independent of the number of spins, characterize the depth of Bell correlations as a function of the interaction range, and determine the critical time at which these correlations emerge. Consequently, we show that scalable quantum resources can be generated with effective short-range interactions for arbitrary large system size $N\gg1$. Finally, we show that the dynamically generated quantum states possessing many-body Bell correlations can be certified with the help of present experimental techniques. 

{\it Spin chains and Bell correlations}---
Let us consider a one-dimensional chain of $N$ spins-$1/2$ described by 
the following Hamiltonian
\begin{align}\label{eq.ham}
  \hat H= \sum_{k,l=1}^{N}J_{kl}\hat\sigma_{z}^{(k)}\hat\sigma_{z}^{(l)},
\end{align}
where $J_{kl}$ is the coupling strength (in units of $\hbar$) of the $k$-th spin interacting with its $l$--th partner, $\hat{\sigma}_z^{(k)}$ are Pauli $z$-operators.
Here, we take the simplest finite-range interaction potential governed by the rectangular function
$J_{kl} =1$ for $0<|k-l|\leqslant r$, and $0$ otherwise. 
{The interaction range} $r$ changes from $r=1$ (nearest-neighbor couplings) to $r=N-1$ (all-to-all couplings), see Fig.~\ref{fig.scheme}. 
A recent experiment~\cite{PhysRevLett.128.113602}
showed a high degree of control over the distance-selective interactions $J_{kl}$. 

Let us start with the  uncorrelated spins polarized along $x$-axis, orthogonal to the $z$-axis distinguished by the Hamiltonian $\hat{H}$, and evolve the system with the time-evolution operator $\hat{U}(\tau) = e^{-i\tau\hat{H}}$, i.e.,
\begin{equation}\label{eq.inix}
  |\psi(\tau)\rangle = \hat{U}(\tau)|1\rangle^{\otimes N}_x,
\end{equation}
and let the interactions correlate spins. Here, $\ket1_{x}$ denotes an eigenstate of the $x$-axis Pauli operator with positive eigenvalue. In the limiting case of the all-to-all interactions, $r=N-1$, system dynamics is equivalent to OAT, where arbitrary entanglement-depth many-body entanglement and Bell-correlated states can be generated \cite{Plodzien2022}. 
There arise natural questions about the critical range $r$, at which quantum features emerge, and about the relation between the range and the pace at which these correlations are generated. 

{A well-suited tool to address these questions is the following correlator
\begin{align}\label{eq.eqn}
\En(\tau)=\modsq{\av{\hat\sigma_+^{(1)}\ldots\hat\sigma_+^{(N)}}}\equiv 2^{Q_N -N},
\end{align}
where the rising operators are taken along the $x$-axis, i.e.,  $\hat\sigma_+^{(k)}=\frac12(\hat\sigma_{y}^{(k)}+i\hat\sigma_{z}^{(k)})$.
This choice of orientation is dictated by the observation that the OAT dynamics generates  GHZ-type superpositions of $\ket{1}_x^{\otimes N}$ and $\ket{-1}_x^{\otimes N}$ states at time $\tau = \pi/4$.

To see how $\En$ relates to the local hidden variable (LHV) theory, replace each $\hat\sigma_{x/y}^{(k)}$ with a $c$-number quantity $\sigma_{x/y}^{(k)}$, that can take binary ($\pm1$) values, depending on the random hidden variable $\lambda$. The locality resides in the assumption that $\sigma$'s depend only on a single label $k$, while realism implies the presence of
a probability distribution $p(\lambda)$ for the hidden variable. Hence, a correlator $\En$ consistent with the LHV theory can be expressed as \begin{align}\label{eq.lhv}
  \mathcal E_N=\modsq{\int\!\! d\lambda\, p(\lambda)\sigma_+^{(1)}(\lambda)\ldots\sigma_+^{(N)}(\lambda)}.
\end{align}
Employing now the Cauchy-Schwarz inequality and the fact that for binary outcomes $\modsq{\sigma_+^{(k)}(\lambda)}=1/2$, we obtain the bound $\mathcal E_N\leqslant 2^{-N}$, which is the $N$-body Bell inequality~\cite{PhysRevLett.126.210506}. Note that in Eq.~\eqref{eq.lhv} we did not assume the $N$ subsystems are described by quantum mechanics. 
If that were the case, the restriction that the quantum spin lies within the Bloch sphere would yield $\modsq{\av{\hat\sigma_+^{(k)}(\lambda)}}\leqslant1/4$, where the mean denotes the trace with the single-qubit density matrix.
Taking all subsystems as quantum-mechanical qubits, we obtain the inequality $\mathcal E_N\leqslant 4^{-N}$. Its violation implies the presence of entanglement
in the system. 
Among these two limiting cases of the structure of the bounds ($\En \leqslant 4^{-N}$ and $\En \leqslant 2^{-N}$), there is a large set of inequalities obtained under assumptions of quantum-mechanical restrictions for some subset of $N$ parties.
Violation of such inequalities implies the EPR steering~\cite{uola2020steering}. Note that the Bell inequality $\En > 2^{-N}$, or equivalently $Q_N>0$ [cf. Eq.~\eqref{eq.eqn}], 
which is the most difficult to violate, ensures entanglement and the EPR steering in the system.

Furthermore, the larger the inequality violation, the stronger correlations {are quantified by the correlator, which are related to the concept of nonlocality depth}.
To see this, take the maximally nonclassical state, the Greenberg-Horne-Zeilinger (GHZ) state
\begin{align}
  \ket{\psi_{\rm GHZ}}=\frac1{\sqrt2}\left(\ket{1}^{\otimes N}_x+\ket{-1}^{\otimes N}_x\right),
\end{align}
which gives the largest value of the Bell correlator~\eqref{eq.eqn}, i.e., $\En=\frac14$, or more conveniently for this purpose, $Q_N=N-2$ [see the r.h.s. of Eq.~\eqref{eq.eqn}]. 
If a single {spin} is not Bell-correlated {in the chain}, i.e., it correlates with the other $N-1$ {spins} according to Eq.~\eqref{eq.lhv}, {we obtain}
 the maximal value of $Q_N=N-3$. Hence, $Q_N>N-3$ implies that the nonlocality encompasses all the spins, and we say that the {\it nonlocality depth}, {denoted with} $\nu$, is $\nu=N$. 
These considerations can be generalized to other integer values of $Q_N$, {and} when
\begin{align}\label{eq:QNrange}
  \nu-3 < Q_N \leqslant \nu-2,
\end{align}
then up to $\nu$ spins are Bell-correlated. 
For further analysis, we introduce the fraction of the Bell correlated particles defined as $\beta=Q_N/N$. 
For more on nonlocality depth and the related depth of entanglement, see Refs.~\cite{zukowski2002bell,cavalcanti2007bell,he2011entanglement,he2010bell,cavalcanti2011unified,spiny.milosz,PhysRevLett.126.210506}.}

{\textit{Universal critical interaction range}}---
\begin{figure}[t!]
  \centering
\includegraphics[width=\linewidth]{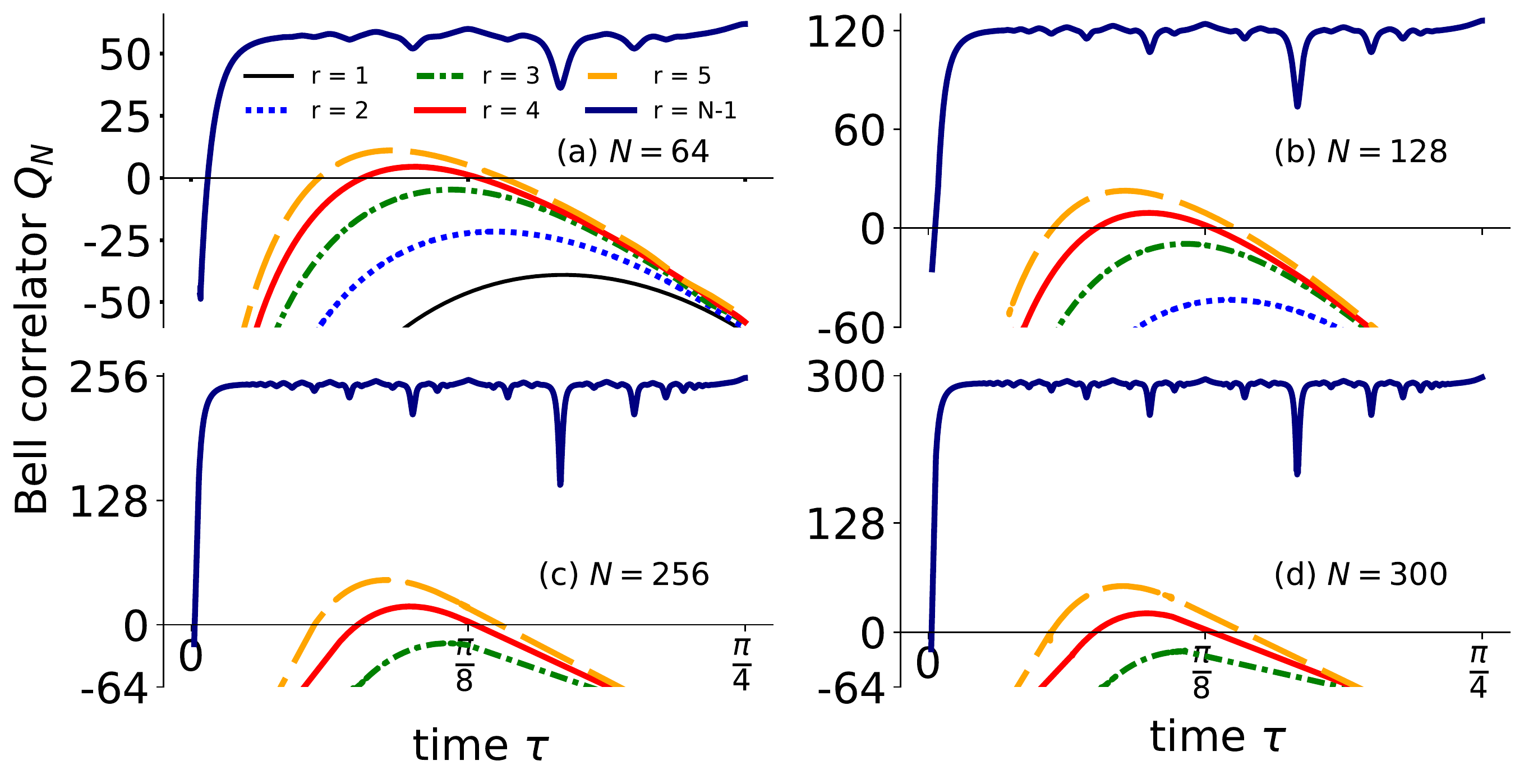}
\includegraphics[width=\linewidth]{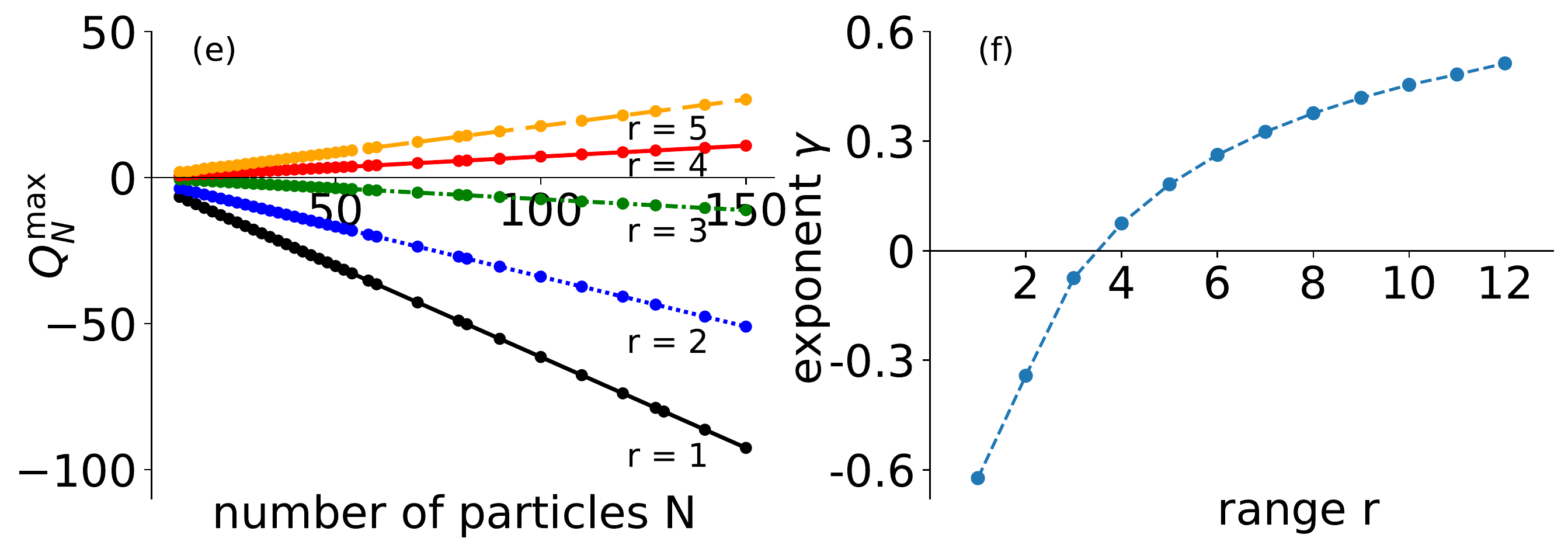}
  \caption{Panels (a)-(d)~The Bell correlator $Q_N (\tau)$ for {$N=64, 128, 256, 300$} spins as a function of time $\tau$ for the interaction range: $r=1$ (thin-solid-black), $r=2$ (dotted-blue) $r=3$ (dashed-dotted-green), 
    $r=4$ (thick-solid-red), $r=5$ (dashed-orange) and $r=N-1$ (solid-dark-blue).   The values $Q_N >0$ mark the region where the many-body Bell inequality is violated.
    (e)~The first maximum of $Q_N$ with respect to time, $Q^\mathrm{max}_N$, as a function of the total number of spins $N$ for various $r \in [1,5]$.
    (f)~The exponent of the Bell correlator in the scaling with $N$, approximated as {$Q_N^\mathrm{max}\approx \gamma N + \mathrm{const}$,} {or, equivalently, $2^N\max_\tau \En(\tau) \propto 2^{\gamma N}$}. 
    The growth of scalable many-body Bell correlations with $N$ is manifested by positive $\gamma$ when $r\geqslant 4$.
    }\label{fig:fig2}
\end{figure}
We {search for} the critical interaction range $r$ that allows the dynamical generation of scalable quantum resources for the arbitrary number of spin $N$, witnessed by Eq.~\eqref{eq.eqn}, during time-evolution Eq.~\eqref{eq.inix}. {In principle, for a given system of size $N$, the minimal value of the range $r$ for which the state exhibits Bell correlations may depend on $N$. Remarkably, as we show below, we find a universal value of this range independent of the length of the chain.}

{First, we note that in} general, the exact numerical calculation of the correlator in Eq.~\eqref{eq.eqn} is exponentially hard as it requires working in the full many-body basis of dimension $2^N$ with access to $N$-body operator $\hat\sigma_+^{(1)}\ldots\hat\sigma_+^{(N)}$. {However,} in the considered scenario, where $[\hat{H}, \hat{\sigma}^{(k)}_z]=0$, the Eq.~\eqref{eq.eqn}{, as we have found,} can be expressed as
\begin{align}\label{eq.ENanal}
  \mathcal E_N(\tau)=\modsqs{ 2^{-N}\sum_{\vec s,\vec s\, '}e^{-i(H_{\vec s}-H_{\vec s'})\tau}s_1\ldots s_N},
\end{align}
where  $H_{\vec s}$ is the $c$-number counterpart of Eq.~\eqref{eq.ham} with operators $\hat\sigma_z^{(k)}$ replaced by $s_k$~\cite{supp}. 
The sums over $\vec s$ and $\vec s\, '$ run over $2^N$ combinations of $s_1,\ldots, s_N$ and $s'_1,\ldots, s'_N$ which take values $\pm 1$. 
The formula in Eq.~\eqref{eq.ENanal} allows for the \textit{exact} calculation of the Bell correlator for the high number of qubits (here up to $N = 300$) since the exponentially growing dimensionality of the Hilbert space is no longer a computational constraint \footnote{The dynamically generated highly entangled states of $N$ spins-$1/2$ in OAT-like protocols have volume law scaling of the entangled entropy $S$, i.e. $S\propto N$. To capture the entangled entropy $S$  the state-of-the-art tensor network methods for studies of one-dimensional spin chains, i.e. Matrix Product States, the bond dimension scales as $S \propto \ln{D}$. As such, to fully capture many-body entangled and Bell correlated states dynamics, including the most entangled many-body GHZ state, the bond dimension should scale exponentially with the number of spins $D \propto e^N$, which is impractical for a large number of spins, here up to $N = 80$. Our analytical approach is free from MPS restrictions and from the course of dimensionality in the considered problem.} (for derivation details see \cite{supp}).

Let us consider first the marginal cases. For the aforementioned $r=1$ case, the Bell correlator is given by
the analytical formula $\mathcal E_N(\tau)=\sin^{N}(\tau)\cos^{3N-4}(\tau)$.
This expression has a maximum ${\rm max}_\tau(\mathcal E_N)\approx 2^{-1.6 N+1}$
at $\tau\simeq\pi/6$, 
which is exponentially smaller than the Bell limit $2^{-N}$. Equivalently, $Q_N = -0.6 N + 1$ and deviation from the Bell limit $Q_N=0$ becomes larger with $N$. Therefore, the nearest-neighbors interactions cannot generate 
many-body Bell correlations in our case. 
The other extreme case is the all-to-all interactions with $r=N-1$, which was considered in detail in~\cite{PhysRevLett.129.250402}. {This case} realizes the one-axis twisting {protocol} and the Bell correlations are present
starting
from $\tau\simeq1.5/N$. The correlator in Eq.~\eqref{eq.eqn} reaches its maximal value $\mathcal E_N=1/4$ at $\tau=\pi/4$, when the $N$-body GHZ 
state is formed \cite{PhysRevA.102.013328, PhysRevLett.129.090403, PhysRevA.107.013311, https://doi.org/10.48550/arxiv.2302.09829}{, which is exponentially larger than the bound in the $N$-body Bell inequality.} 

In Fig.~\ref{fig:fig2} in panels (a)---(d) we present time evolution of the correlator $Q_N(\tau)$, calculated using Eq.~\eqref{eq.ENanal}, for different intermediate ranges $1\leqslant r \leqslant 5$ and for all-to-all couplings $r=N-1$ for various spin numbers $N = 64$ (panel a), 128 (b), 256 (c) and 300 (d). We observe that the correlator only breaks the Bell limit $Q_N>0$ when $r\geqslant4$ even for large $N$. This is the first indication of the presence of a critical range. However, it must be verified if the value of $r=4$ is universal, i.e., independent of the system size. To this end, we focus on
\begin{align}\label{maxQ}
  Q_N^{\rm max}\equiv\max_{\tau}Q_N(\tau),
\end{align}
i.e., maximized $Q_N$ with respect to $\tau$ calculated for different $N$'s.
In Fig.~\ref{fig:fig2}(e) we show $Q_N^{\rm max}$, as a function of the number of spins $N$ and for various values of $1\leqslant r\leqslant 5$. For a given $r$, these maximal values lie on a straight line determining an exponent $\gamma$ in the exponential scaling of  $\En\propto 2^{(\gamma -1)N}$. 
In Fig.~\ref{fig:fig2}(f), we present the value of $\gamma$ as a function of the interaction range. For $r\geqslant 4$ we find positive exponents indicating that the degree of the Bell inequality violation becomes higher with growing $N$ in contrast to the case of $r\leqslant 3$. Therefore, the change in the sign of $\gamma$ indicates qualitatively different scaling regimes of the Bell correlations.

The observed behaviour of $Q_N^{\rm max}$ and $\gamma(r) > 0$ for $r\geqslant 4$ demonstrates that $r=4$ is indeed a 
{\it critical range}, at which the Bell correlations are detected. Crucially, $Q_N^{\rm max}$ increases with growing $N$, 
indicating scalable Bell correlations in this system~\cite{PhysRevLett.126.210506}. 
Remarkably, although $r=4$ in the limit $N\gg1$  is a short-range interaction encompassing an intensive number of neighboring spins, the system still exhibits an increasing degree of violation of the Bell inequality {with increasing the system size}. The invariance of the critical range with respect to the number of spins proves {its \textit{universality}}.

Qualitatively, the strengthening of Bell correlations with growing $r$ can be explained as follows. 
To generate the GHZ-like coherence, which is the witness of Bell correlations [see Eq.~\eqref{eq.eqn}] one needs to flip all the spins, hence, to act with an $N$-body operator 
$\hat{\mathcal B}_N=\hat\sigma_z^{(1)}\ldots\hat\sigma_z^{(N)}$ on the input state,
since $\hat\sigma_z\ket1_x=\ket{-1}_x$. Note that the evolution operator
coupling any pair of spins is
\begin{align}\label{eq.evol}
  e^{-i\tau \hat\sigma_z^{(k)} \hat\sigma_z^{(l)}}=\cos(\tau) - i\sin(\tau) \hat\sigma_z^{(k)} \hat\sigma_z^{(l)}.
\end{align}
As $r$ grows, each spin couples to more neighbors, and the number of possible combinations of interacting terms giving the operator $\mathcal B_N$ grows exponentially, increasing, in consequence,
$\En(\tau)$. This growth is, however, balanced by the exponential decay of the amplitude of the initial state. The eventual observation of the Bell inequality violation is, thus, the effect of the competition between the two indicated mechanisms. We formalize these intuitions with the help of the developed spin-inversion asymptotic expansion theory; for quantitative explanations using diagrammatic approach~\footnote{For intermediate values of $r$, 
in our open-access repository~\cite{code_repo}, we provide examples of the time evolution of $\mathcal E_N(\tau)$ 
obtained analytically for concrete values of $N$ and $r$.} and the details on the asymptotic theory, see~\cite{supp}.

{We note here that identifying relevant mechanisms helps} clarify the role of the boundary conditions. The periodic ones provide some additional coupling between the spins. Indeed, numerical results
confirm that in {such a} case, $\En(\tau)$ {is, in general, larger as compared} to the open boundary conditions considered here. Our decision to focus on the open ones is motivated by the geometry of
most of the experimental setups.
\begin{figure}[t!]
  \centering
  \includegraphics[width=\linewidth]{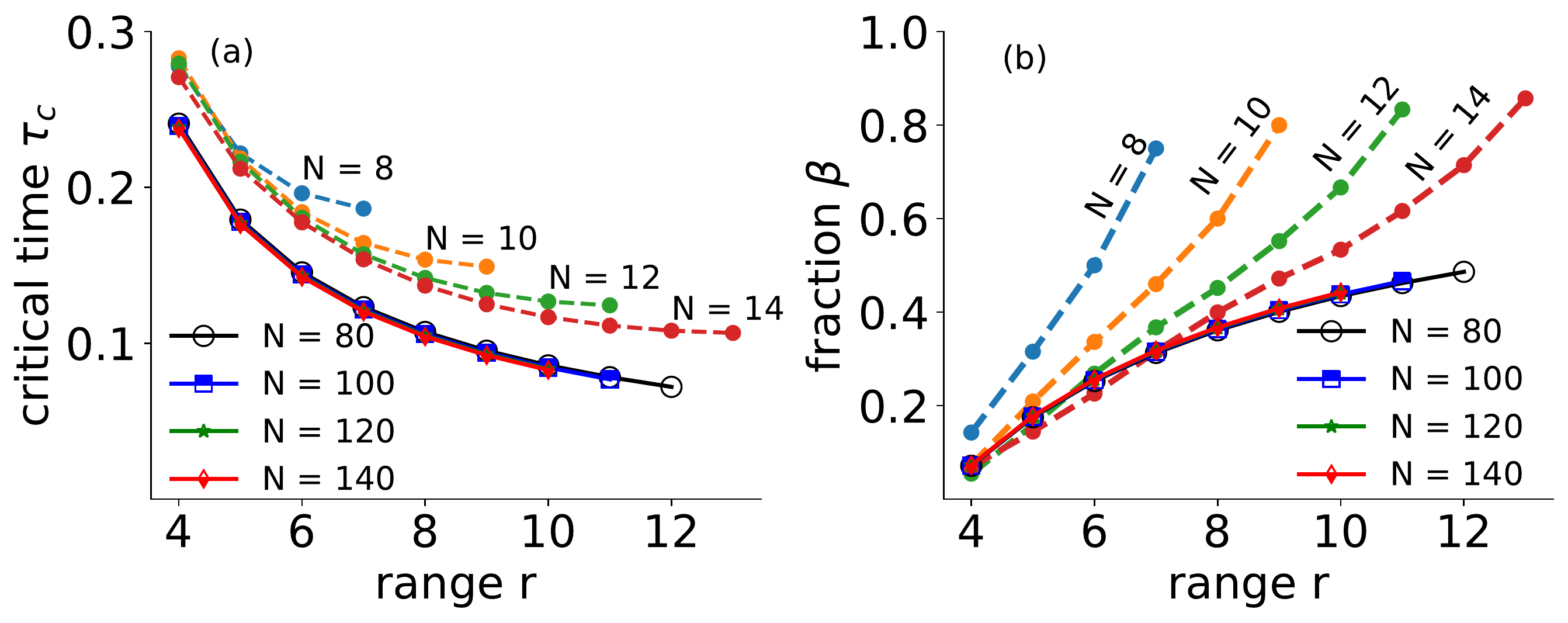}
  \caption{(a)~The critical time $\tau_c$ at which the Bell correlator passes the Bell limit, $Q_N=0$, as a function of interaction range $r$. 
The dashed curves correspond to the limit $r\lesssim N$, while solid curves correspond to the limit $r\ll N$. The latter collapses onto the same line. (b)~The maximal fraction of correlated spins $\beta = \nu/N$ as a function of range $r$, cf. Eq.~\eqref{eq:QNrange}. 
}
\label{fig:fig3}
\end{figure}

{\it Critical time}---
{As evident from Fig.~\ref{fig:fig2}(a)---(d), the interaction range $r$ determines the time, denoted with $\tau_c$, at which Bell correlations emerge. This critical time, at which the correlator $Q_N(\tau)$ crosses the Bell limit, depends on $r$, and the generation of correlations accelerates for larger $r$'s.}
This dependence can be extracted from  Eq.~\eqref{eq.ENanal} {as follows}. For an arbitrary internal spin coupled with $2r$ neighbours, the sum over $s_k=\pm1$ vanishes unless the corresponding phase-term $e^{-i\tau s_k(s_{k-r}+\ldots +s_{k+r})}$ oscillates quickly enough.
The sum in the parenthesis is at {most} equal to $2r$, hence the phase factor will vary significantly between $s_k=-1$ and $s_k=1$ if $\tau\cdot2r\gtrsim1$. Thus, the Bell correlator becomes significantly
non-zero if $\tau_c\gtrsim a/r$, where $a$ is some constant. {The scaling of $\tau_c$, which is inversely proportional to the interaction range, is} confirmed by exact solution of the dynamics generated by Eq.~\eqref{eq.ham}. 
In Fig.~\ref{fig:fig3}(a), we identify the
critical time $\tau_c$ when $Q_N$ {surpasses} the Bell limit for different values of large $N$=80, 100, 120, 140, for which $r\ll N$ and small $N$=8, 10, 12, 14, where $r\lesssim N$. The observed behavior shows that for large systems, $\tau_c$ becomes independent of the system size. By changing the range $r$ from the critical value $r=4$ to $r=N=14$ we recover the {scaling} $\tau_c\propto 1/N$
{characteristic to the all-to-all coupling case.}

{\it Fraction of Bell correlations spins}--- 
In Fig.~\ref{fig:fig3}(b), 
we present the dependence of the fraction of Bell correlated spins $\beta=Q_N/N$ as a function of the range $r$, deduced from $Q_N^{\rm max}$, see Eq.~\eqref{eq:QNrange}. 
In the limit of $N\gg r$, the results follow the universal curve.
This can be understood as follows. Using the asymptotic exponential scaling for large $N$, i.e., $Q_N \approx \gamma(r)N$ [see the discussion below Eq.~\eqref{maxQ}], we can estimate the nonlocality depth by $\nu_N \approx Q_N$ up to corrections on the order of $1/N$. Then, the fraction $\beta \approx \gamma(r)$, which shows that for large $N$, $\beta$ becomes independent of $N$ as is observed in the figure. This relation uncovers the physical role of the exponent of $\mathcal E_N$ as a concentration of Bell correlated particles. Therefore, we find the range not only controls the critical time $\tau_c$, but also the number of Bell-correlated clusters in the system.

\begin{figure}[t]
\centering
\includegraphics[width=\linewidth]{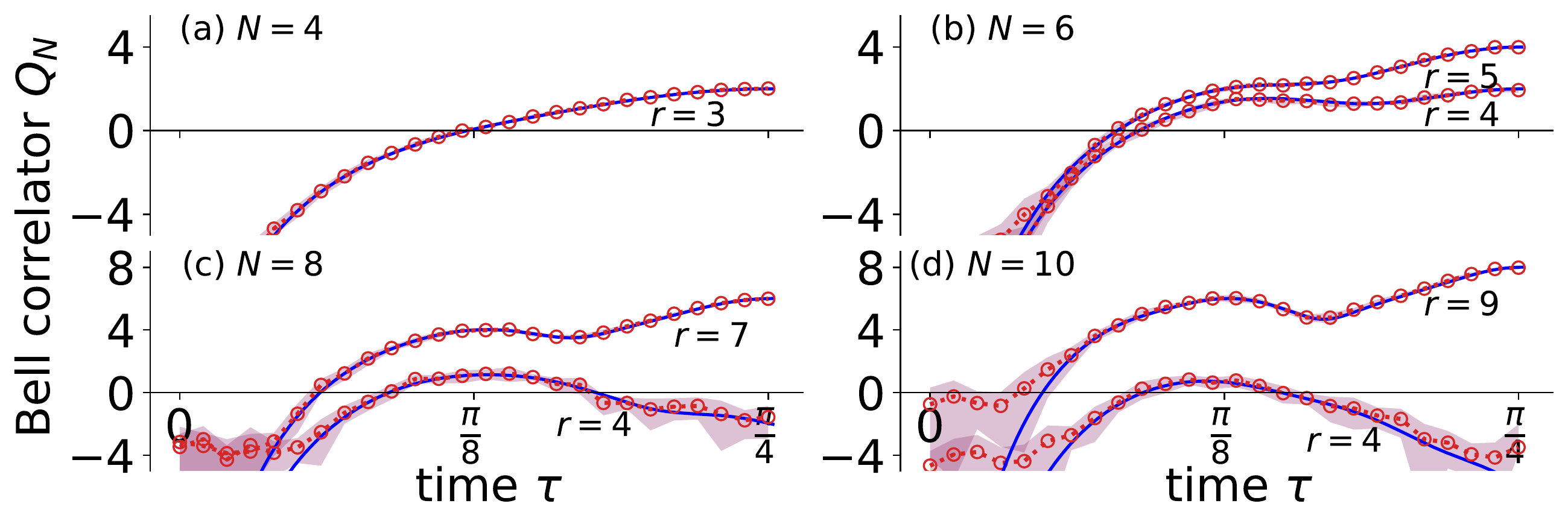}
    \caption{
    The dynamics of the estimated many-body Bell correlator $Q_N$ for (a) $N=4$, (b) $N = 6$, (c) $N = 8$, and (d) $N = 10$ spins with  
    interaction range $r = 4$ (lower curves), and all-to-all connections  $r=N-1$ (upper curves).
    Solid blue lines present exact results, while red circles represent the value of the reconstructed many-body Bell correlator from the classical shadows tomography. Standard deviation is marked as a shaded area.
    }\label{fig:fig4}
\end{figure}
{\it Many-body Bell correlation certification}--- Measuring high-order correlation functions is challenging.
Recent advances in the control of many-body systems allowed for measurements of correlation functions up to the $6$-th order \cite{Dall2013}, $2$-nd  R\'enyi entropy  for $N=4$ \cite{Islam2015}, and $N=5$ particles \cite{Linke2018}  via extraction characteristic of a quantum state using a controlled-swap gate acting on two copies of the state \cite{Eckert2002}, while for $N=10$ particles \cite{Brydges2019} using the randomized measurements technique \cite{Vermersch2018,Elben2019,Elben2020,Elben2020_PRL,Rath2021,Elben2022}.
The many-body correlations considered here, however, can be certified by inspecting only one element of the density matrix, which couples the state with all spins up with the state with all spins down in the $x$-basis, cf. Eq.~\eqref{eq.eqn}. As such, the problem of many-body Bell correlations measurement can be cast as a quantum state tomography task \cite{PhysRevLett.74.4101,PhysRevLett.83.3103,PhysRevLett.92.220402, Haffner2005, 5714248,Toth2010,Moroder_2012,Cramer2010,PhysRevLett.111.020401,Lanyon2017}, which recently has been enhanced by deep neural networks techniques \cite{adri1,Pan2022,koutny2022PRA,Shahnawaz2021,ma2023attentionbased,palmieri2023enhancing, dawid2022modern}.

To this end, we simulate tomographical reconstruction of the density matrix with classical shadows tomography (CST) ~\cite{10.1145/3313276.3316378,Huang2020,altepeter20044, 7956181, 10.1145/2897518.2897544,Koh2022classicalshadows,aaronson2018shadow,PhysRevLett.125.200501, Huang2020, PhysRevLett.127.030503, Elben2023}, which {was employed for} trapped ions \cite{mcginley2022shadow,PhysRevX.13.011049,doi:10.1126/science.1208001}, and Rydberg atom arrays \cite{Notarnicola_2023}. We prepared $10$ reconstructions of the 
target density matrix $\hat{\varrho}(\tau)$ at a given time $\tau$, where
each reconstruction consists of $M$ classical shadows. From each reconstructed $\hat{\varrho}(\tau)$, we extract the Bell correlator, and based on the generated collection, we estimated the mean and standard deviation of the reconstruction ${\cal E}^*(\tau) \equiv 2^{Q_N^*(\tau)-N}$~\cite{supp}. 
In Fig.~\ref{fig:fig4}, we demonstrate that Bell correlations quantified by $Q_N(\tau)$ (solid blue lines) can be successfully certified by the tomographically reconstructed mean value $Q_N^*(\tau)$ (red circles) with standard deviation (shaded areas) for $N = 4,6,8,10$ spins with the interaction range $r = 4$, and the all-to-all couplings when $r=N-1$. We prepared $M=10^5 N$ classical shadow for $N = 10$ and $r = 4$, and $M = 10^4 N$ otherwise.

An alternative method to full quantum state tomography relies on the {concept of} multiple quantum coherences (MQC), which provide extensive information about the structure of a many-body state and allows for the relation of the quantum correlators with other physical quantities like the out-of-time-order correlations~\cite{garttner2017measuring,PhysRevLett.120.040402,supp}. In SM we provide details how $\En$ can be determined by means of high-order multiple-quantum intensity in MQC which is feasible experimentally.

{\it Discussion and Conclusions}---
In this work, we showed that scalable many-body entanglement and Bell correlations can be generated with the Ising-type Hamiltonian with finite-range two-body interactions, in {contrast} to the broadly considered all-to-all couplings present in the standard one-axis twisting protocols. {We found a critical interaction range that is} independent of the system size, which means that for large system size $N\gg1$ interactions are {short-ranged}. Yet, the degree of violating the many-body Bell inequality grows with the number of spins. 
Moreover, the studied quantum resources can be experimentally certified with the classical shadows quantum state tomography and multiple-quantum coherences measurements. 
 
Our study addresses the constant need for protocols to generate and certificate $N$-body entanglement in quantum circuits. First, our results have important consequences for future digital quantum computers operating on large numbers of qubits, aiming to generate scalable quantum resources dynamically. In a single trotterization step of the time-evolution in the {one-axis twisting} protocol, there is a need for the all-to-all connections between qubits; thus, the number of two-qubit entangling gates scales quadratically with the number of qubits \cite{PhysRevLett.119.180511, doi:10.1126/science.aay0600, doi:10.1126/sciadv.aba4935, Li2018}, while for the finite-critical-range interactions considered here, it scales linearly {which is a significant reduction of the problem complexity}. Next, our analytical results allow us to extract many-body quantum correlations, in principle, for an arbitrary number of qubits, which can be used to certify and benchmark quantum computers operating on hundreds of qubits \cite{wurtz2023aquila,Kim2023}.

{\it Author contributions}---
M.P. indicated the existence of the critical interaction range, performed many-body simulations, analytical calculations in Ref.~\cite{code_repo}, 
and quantum state tomography results. J.Ch. prepared  analytical expressions for the Bell correlator, critical range, and critical time, and performed many-body calculations. 
T.W. developed the spin-inversion asymptotic expansion theory and performed many-body calculations. 
All the authors contributed to discussing the results and the manuscript preparation and revision.

{\it Ackonwledgements}---
We acknowledge the discussion with Arghavan Safavi-Naini, Rene Gerritsma, Rainer Blatt and Tomasso Roscilde. J.Ch. was funded by the National Science Centre, Poland, within the QuantERA II Programme that has received funding from the European Union’s Horizon 2020 research and innovation programme under Grant Agreement No 101017733, Project No. 2021/03/Y/ST2/00195.
E.W. acknowledges support from the DAINA project of the Polish National Science Center DEC-2020/38/L/ST2/00375.
This research is part of the project No. 2021/43/P/ST2/02911 co-funded by the National Science Centre and the European Union’s Horizon 2020 research and innovation programme under the Marie Skłodowska-Curie grant agreement no. 945339. For the purpose of Open Access, the author has applied a CC-BY public copyright licence to any Author Accepted Manuscript (AAM) version arising from this submission.
ICFO group acknowledges support from: ERC AdG NOQIA; Ministerio de Ciencia y Innovation Agencia Estatal de Investigaciones (PGC2018-097027-B-I00/10.13039/501100011033, CEX2019-000910-S/10.13039/501100011033, Plan National FIDEUA PID2019-106901GB-I00, FPI, QUANTERA MAQS PCI2019-111828-2, QUANTERA DYNAMITE PCI2022-132919, Proyectos de I+D+I “Retos Colaboración” QUSPIN RTC2019-007196-7); MICIIN with funding from European Union NextGenerationEU(PRTR-C17.I1) and by Generalitat de Catalunya; Fundació Cellex; Fundació Mir-Puig; Generalitat de Catalunya (European Social Fund FEDER and CERCA program, AGAUR Grant No. 2021 SGR 01452, QuantumCAT \ U16-011424, co-funded by ERDF Operational Program of Catalonia 2014-2020); Barcelona Supercomputing Center MareNostrum (FI-2022-1-0042); EU (PASQuanS2.1, 101113690); EU Horizon 2020 FET-OPEN OPTOlogic (Grant No 899794); EU Horizon Europe Program (Grant Agreement 101080086 — NeQST), National Science Centre, Poland (Symfonia Grant No. 2016/20/W/ST4/00314); ICFO Internal “QuantumGaudi” project; European Union’s Horizon 2020 research and innovation program under the Marie-Skłodowska-Curie grant agreement No 101029393 (STREDCH). 
M.P. acknowledges the support of the Polish National Agency for Academic Exchange, the Bekker programme no:
PPN/BEK/2020/1/00317.
Views and opinions expressed are, however, those of the author(s) only and do not necessarily reflect those of the European Union, European Commission, European Climate, Infrastructure and Environment Executive Agency (CINEA), nor any other granting authority. Neither the European Union nor any granting authority can be held responsible for them.

\bibliography{bibl}

\clearpage
\setcounter{equation}{0}
\setcounter{figure}{0}
\setcounter{table}{0}

\makeatletter
\renewcommand{\theequation}{S\arabic{equation}}
\renewcommand{\thefigure}{S\arabic{figure}}
 
\clearpage
\onecolumngrid
\begin{center}
    {\large\bf Supplemental Material for\\[2mm] ``Generation of scalable many-body Bell correlations in spin chains with short-range two-body interactions''}\\[3mm]
    Marcin P\l{}odzie\'n,$^1$ Tomasz Wasak,$^2$ Emilia Witkowska,$^3$
    Maciej Lewenstein,$^{1,4}$ and Jan Chwede\'nczuk$^5$\\[1mm]
{\small 
$^1${\it ICFO - Institut de Ci\`encies Fot\`oniques, The Barcelona Institute of Science and Technology, 08860 Castelldefels (Barcelona), Spain}\\
$^2${\it Institute of Physics, Faculty of Physics, Astronomy and Informatics, Nicolaus Copernicus University
in Toru\'n, Grudzi\c{a}dzka 5, 87-100 Toru\'n, Poland }\\
$^3${\it Institute of Physics PAS, Aleja Lotnikow 32/46, 02-668 Warszawa, Poland}\\
$^4${\it ICREA, Pg. Lluis Companys 23, ES-08010 Barcelona, Spain}\\
$^5${\it Faculty of Physics, University of Warsaw, ul. Pasteura 5, PL-02-093 Warsaw, Poland}
}
\end{center}
\setcounter{page}{1}
\vspace*{4mm}

\begin{bibunit}
\twocolumngrid

\section{Derivation of formula from Eq.~(7)}\label{app.der.corr}

In this section, we derive Eq.~(7) from the main text. We begin by invoking a general expression for the density matrix of $N$ qubits which is 
\begin{align}\label{eq.app.input}
  \hat\varrho_{\rm in}=\sum_{\vec s,\vec s'}\varrho_{\vec s,\vec s'}\ketbra{\vec s}{\vec s'}.
\end{align}
The double sum over $\vec s$ and $\vec s'$ is a shortened notation for $2N$ sums over $s_1,\ldots, s_N=\pm1$ and $s'_1,\ldots, s'_N=\pm1$. 
The considered Hamiltonian 
\begin{align}
  \hat H=\sum_{k,l=1}^N J_{kl}\hat\sigma_z^{(k)}\hat\sigma_z^{(l)},
\end{align}
where $J_{kl} = 1$ for $0<|k-l|\leqslant r$, and $0$ otherwise,
 can be expressed in the following form
\begin{align}\label{eq.app.ham}
  \hat H =\sum_{r=1}^{N-1} \sum_{k=1}^{N-r}\hat\sigma_z^{(k)}\hat\sigma_z^{(k+r)},
\end{align}
Since the Hamiltonian contains only $\hat{\sigma}_z^{(l)}$ operators, it is natural to use as the basis the product of eigenstates of $N$ $z$-axis Pauli matrices. Hence the action of the evolution operator on each such ket replaces all Pauli operators in
Eq.~\eqref{eq.app.ham} with a corresponding set of $\vec s$ numbers, i.e., 
\begin{align}
  e^{-i\hat H\tau}\ket{\vec s}&=e^{-i\tau\sum_{r=1}^{N-1}\sum_{k=1}^{N-r}\hat\sigma_z^{(k)}\hat\sigma_z^{(k+r)}}\ket{\vec s}_z=\\
  &=e^{-i\tau\sum_{r=1}^{N-1}\sum_{k=1}^{N-r}s_ks_{k+r}}\ket{\vec s}_z\equiv e^{-i\hat H_{\vec s}\tau}\ket{\vec s}_z.\nonumber
\end{align}
Therefore, the output state becomes
\begin{align}\label{eq.app.out}
  \hat\varrho_{\rm out}(\tau)=e^{-i\hat H\tau}\hat\varrho_{\rm in}e^{i\hat H\tau}=\sum_{\vec s,\vec s'}e^{-i(\hat H_{\vec s}-\hat H_{\vec s'})\tau}\varrho_{\vec s,\vec s'}\ketbra{\vec s}{\vec s'}_z
\end{align}
Since the correlator $\mathcal E_N$ is calculated with the product of $N$ operators rising the spin projection along the $x$-axis, it is convenient to change the basis using
\begin{align}
  \ket{s_k}_z=\frac1{\sqrt2}(\ket{{+1}_k}+s_k\ket{{-1}_k}),
\end{align}
where the absence of the subscript $z$ indicates that we are working in the eigenbasis of $x$-axis Pauli operators. The above expression can be expressed in a compact way as follows
\begin{align}
  \ket{s_k}=\frac1{\sqrt2}\sum_{m_k=\pm1}s_k^{\frac{1-m_k}2}\ket{m_k}.
\end{align}
Substituting this result into Eq.~\eqref{eq.app.out} we obtain
\begin{align}
  \hat\varrho_{\rm out}(\tau)&=\frac1{2^N}\sum_{\vec s,\vec s'}e^{-i(\hat H_{\vec s}-\hat H_{\vec s'})\tau}\varrho_{\vec s,\vec s'}\nonumber\times\\
  &\times\sum_{\vec m, \vec m'}\ketbra{\vec m}{\vec m'}
  \prod_{k=1}^Ns_k^{\frac{1-m_k}2}s_k^{'\frac{1-m'_k}2}.
\end{align}
The correlator couples the two extreme elements of the density matrix: the one where all spins are up: $\ket{+1}^{\otimes N}$ with that where all are down
$\ket{-1}^{\otimes N}$. Hence it can be expressed by a single element of $ \hat\varrho_{\rm out}(\tau)$, namely the coherence term between these two elements. This
corresponds to setting all $m$'s to $-1$ and all $(m')$'s to $+1$, giving
\begin{align}\label{eq.app.corr}
  \mathcal E_N&=\modsqs{\varrho_{+1^{\otimes N},-1^{\otimes N}}}=\\
  &=\modsqs{\frac1{2^N}\sum_{\vec s,\vec s'}e^{-i(\hat H_{\vec s}-\hat H_{\vec s'})\tau}\varrho_{\vec s,\vec s'}s_1\ldots s_k}\nonumber
\end{align}
as invoked in Eq.~(\ref{eq.ENanal}) of the main text.

\section{Diagrammatic spin-inversion asymptotic expansion theory}\label{sec:spinflip}

To capture the physical mechanism of the correlations' enhancement beyond the universal critical range $r=4$, we resort to an asymptotic diagrammatic expansion of the many-body state coefficients in the number of spin-inversions throughout the evolution. The proposed method  provides information about the time dependence of $\En$, enabling the estimation of the critical time and indicating the origin of enhanced correlation at the critical range $r=4$. 

We consider first the part of the 
Hamiltonian from Eq.~\eqref{eq.ham} which couples a single pair, i.e., $\hat H_{kl}=J_{kl}\hat\sigma_{z}^{(k)}\hat\sigma_{z}^{(l)}$. Then, under the action of $\hat H_{kl}$, we obtain
\begin{align} \label{f}
  e^{-i\tau \hat H_{kl}}\ket{1_k}\ket{1_l} = \hat f_{kl} \ket{1_k}\ket{1_l},
\end{align}
where $\hat f_{kl} = \cos(J_{kl}\tau) - i\sin(J_{kl}\tau) \hat\sigma_z^{(k)} \hat\sigma_z^{(l)}$. The action of the second term in the operator $\hat f_{kl}$ inverts the spins due to the property $\hat\sigma_z^{(k)} \ket{1_k} = \ket{{-1}_k}$ 
unless the distance $d=|k-l|$ exceeds the range $r$. Since the evolution of the state is given by a product of the terms from Eq.~\eqref{f},
\begin{equation}\label{eq.intime}
  \ket{\psi(\tau)} = \prod_{k,l}{}\hat{f}_{kl}  \ket{\psi_\mathrm{in}},
\end{equation}
we can represent the final state diagrammatically as a sequence of dots (each corresponding to the initial state of the individual spin) and lines connecting pairs of dots. 
To each line connecting a pair we assign either the amplitude $\cos(\tau)$ and an unchanged state of the pair, or the amplitude $- i\sin(\tau)$ to a pair with inverted spins. 
The final state is the sum over all possible assignments to all the lines and all possible ways of connecting the pairs of spins that are compatible with the distance of the interaction.
In such a case, the nontrivial lines can connect
only points that are not further apart than the distance $r$, and there are in total $K \equiv r(2N -r-1)/2$ of such lines. 
We now apply this diagrammatic method to calculate the $\En$ at short times and identify at which $r$ the correlator crosses the Bell limit.

\subsubsection{Diagrammatic approach}
\label{app:diag}

Here, we describe the diagrammatic method allowing for physical insights regarding the origin of the critical interaction range $r = 4$.
We begin by analyzing the   number of possibilities of forming the diagrams $P_r$  for small values of $r$. We show that the number of diagram classes in the case of $r=4$ exceeds those of $r=3$. As a result, new additional diagrams appear for $r=4$, which describes a correlation that spreads over the whole chain and thus increases the exponent of $P_4(N)$. 
\begin{figure}[t]
\centering
  \includegraphics[width=\linewidth]{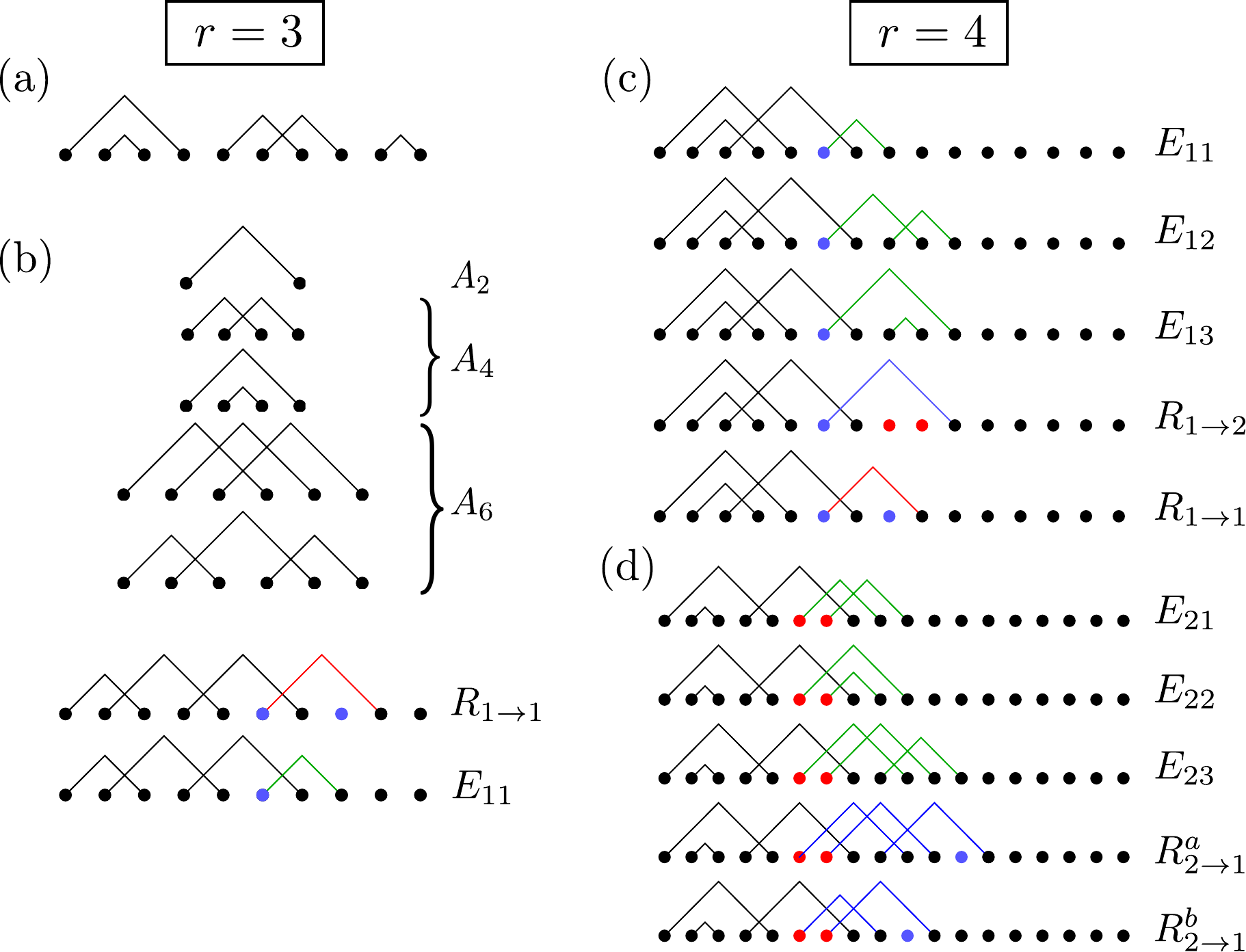}
  \caption{(a) An example of diagrams for $r=3$ and $N=10$. Three disconnected cluster diagrams are visible. 
    (b) The classes of clusters including 2, 4 and 6 spins ($A_{2,4,6}$) for $r=3$. $R^{a/b}_{x \to y}$ and $E_{xy}$ with $x,y=1,2,3$ mark extended and closing cluster diagrams as explained in the main text. (c,d) The classes of cluster diagrams with one (two) unconnected spins are marked
    with blue (red) points. The green lines close the cluster, the blue lines exchange the classes between one and two unconnected last spins. The red line extends the diagram within the same class.
    }
    \label{fig:fig7}
\end{figure}

To proceed, we first investigate the case $r=3$, see Fig.~\ref{fig:fig7}(a). As can be seen from the
diagram which is an example for $N=10$, there are clusters or blocks of spins, between which there are no lines. The positions of such different clusters can be exchanged leading to an exponential scaling with $N$ due to many permutations when $N$ grows. These clusters including $k=2$, 4 and 6 spins, which form classes $A_{2,4,6}$, see Fig.~\ref{fig:fig7}(b), can appear anywhere in the chain. 

For comparison, in the case $r=1$ only a single class $A_2$ from Fig.~\ref{fig:fig7}(b) contributes, which contains just a single element. As a result, there are $(N/2)!$ such permutations built from the same diagram, and thus $P_1 = (N/2)! / (N/2)! =1$. In the case, $r=2$, in addition to the class $A_2$, the first diagram from the class $A_4$ from Fig.~\ref{fig:fig7}(b) contributes. Here, we can perform the summation analytically and, after straightforward calculations, we obtain $P_2(N) = F_{N/2+1}$,
where $F_n$ is the $n$-th Fibonacci number.
For large $N$, we have $P_2 \propto 
e^{\ln\phi\, N/2}$, which confirms the exponential scaling with the exponent $(\ln\phi)/2 \approx 0.241$ determined by the golden ratio $\phi$.

Importantly, in the case of $r=3$, there is a single cluster diagram, the class denoted by $R_{1\to1}$ in Fig.~\ref{fig:fig7}(b), which can be extended up to arbitrary $k\leqslant N$ spins by adding a single line shown in red starting from the blue point. The notation $R_{p\to q}$ is meant to denote a recursive type of diagram that brings $p$ open points to $q$ open points in the cluster.
This extended cluster diagram can be terminated at any point by adding the green line as shown in the diagram denoted by $E_{11}$.
Interestingly, such a cluster can include all spins. The appearance of such extended diagrams increases the exponent of $P_r$ to $0.427$, which we found by a simple fit, and so approximately by a factor of 2 compared to the case $r=2$. This increase, however, is insufficient to reach the Bell limit.

For the case $r=4$, the situation changes qualitatively. Similarly to the $r=3$ case, small cluster diagrams appear in the expansion, but now the class of extended diagrams is much larger. In Fig.~\ref{fig:fig7}(b), we show examples of diagrams which end with one unbound spin (blue). Not only the cluster can be closed by adding
one of the green lines from classes $E_{1,j}$ with $j=1,2,3$, but also it can be further extended by using the red line from class $R_{1\to1}$. Alternatively, the diagram
can be converted to a cluster ending with two spins, as in class $R_{1\to2}$. Next, such diagrams can be closed (using green lines from classes $E_{2,1/2/3}$ in panel
(c)), or converted to a diagram with one unbound spin (as in panel (b)) by using blue lines from class $R^{a/b}_{2\to1}$. As a consequence, by fitting an exponential, we find $0.563$ for the exponent $P_4$. 
For completeness, we mention that we found the exponent $0.670$ for $r=5$.

The increase of the exponent of $P_4(N)$, relative to the $r=3$ case, is sufficient to surpass the Bell limit. 
Such long-range cluster diagrams have a much larger contribution in the case $r=4$ than for $r=3$. For instance, for $N=10$, there are 15 different cluster diagrams embodying all the $N$ spins for $r=4$ 
compared to a single diagram for $r=3$.  As a result, the rate of increasing $P_r(N)$ as a function of $N$ is larger with increasing $r$ and leads to crossing the Bell correlations limit when $r=3$ 
is increased to $r=4$. 

\subsubsection{Expansion of correlator $\En$}

The correlator $\En(\tau)$ from Eq.~\eqref{eq.eqn} can be conveniently rewritten in the form
\begin{equation} \label{En_cc}
  \En(\tau) = |C_+(\tau)|^2 \, |C_-(\tau)|^2,
\end{equation}
for the pure state from Eq.~(\ref{eq.intime}), where $C_+$ $(C_-)$ is the amplitude of finding all spins up (down) along the $x$-axis. Therefore, in order to calculate the amplitude $C_-$ we need to count all the processes that lead to the spin-inversion of all $N$ spins, and in order to obtain $C_+$ we have to take into account all the processes that resulted in no net spin-inversion of the initial state. 
Our approach is to consider only those diagrams in which each spin inverted the fewest number of repetitions -- which is justified for short times.

We first determine $C_-(\tau)$, focusing on the dominating process where all the spins have inverted only once. 
Diagrammatically, each pair of points is connected only once with a line with the amplitude $-i \sin{\tau}$, and there are $N/2$ such lines, while all the remaining $K-N/2$ lines have amplitudes
$\cos\tau$.  Therefore, the coefficient $C_-$ is given by
\begin{equation}\label{Cd_approx}
  C_- \approx P_r(N) \big(\cos\tau\big)^{K-N/2} \big(- i \sin\tau\big)^{N/2},
\end{equation}
where $P_r(N)$ denotes the number of possibilities of forming the diagrams. 

A similar expansion holds for $C_+$, but now we count the diagrams with the least number of lines with the amplitude $-i \sin{\tau}$ that result in no net spin inversion. 
In the zeroth order, no spin inversions occur resulting in the scaling of $\cos^K\tau$. In the first order, such diagrams contain three lines where the three spins $i$, $j$, and $k$ are mutually connected and subject to the constraint that all the distances between the pairs fall within the range of the interaction. By the symbol $R_r(N)$, we denote the number of such diagrams for the given range $r$ and the number of spins $N$. We thus arrive at the following expression
\begin{equation}\label{C1_approx}
  C_+ \approx \big(\cos\tau\big)^{K} + R_r(N) \big(\cos\tau\big)^{K-3} \big(- i \sin\tau\big)^{3}.
\end{equation}
For $r=1$ we only have a single diagram with only nearest-neighbours being connected, and, thus, $P_{r=1} = 1$ and $R_{r=1} = 0$. As a result, we recover the exact formula cited in 
the main text, namely
\begin{align} \label{En_nn}
    \mathcal E_N(\tau)=\sin^{N}(\tau)\cos^{3N-4}(\tau).
\end{align}
For small values of $r$ or $N$, the actual numbers for $P_r$ and $R_r$ can be found by counting the number of permutations or from analytical solutions~\cite{code_repo}. In the large $N$-limit, $R_r$ scales linearly with $N$, and $P_r$ scales exponentially with an $r$-dependent exponent.
This exponential growth of $P_r$ with a sufficiently large exponent is responsible for surpassing the Bell {limit} and observation of nonlocal correlations.

\begin{figure}[]
  \centering
  \includegraphics[width=\linewidth]{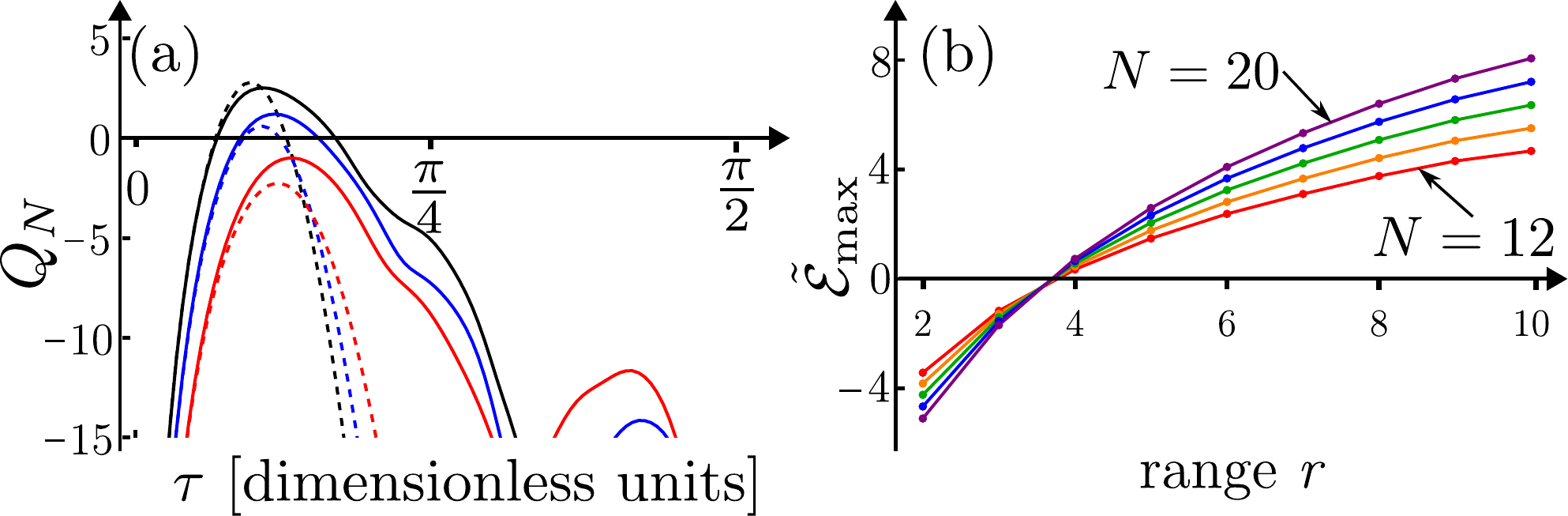}
  \caption{
      (a) The comparison of exact (solid) and approximate (dashed) numerics for $N=16$ and $r = 3$ (red), 4 (blue), 5 (black). The parameters used are $(P_r,R_r) = (491, 40)$ for $r=3$, (3116, 76) for $r=4$ and (12483, 120) for $r=5$.
    (b) The behaviour of $\tilde{\mathcal{E}}_\mathrm{max}$ as a function of the range $r$ for various values of $N = 12$, 14, 16, 18, 20. The solid line is a guide to the eye. The value of $\tilde{\mathcal{E}}_\mathrm{max}$ is positive for $r=4$ which signals the existence of $\tau_\mathrm{crit}$ for the Bell inequality violation as given by Eq.~\eqref{tcrit.approx}. 
    }\label{fig:fig5}
\end{figure}

In Fig.~\ref{fig:fig5}(a), we compare our asymptotic expansion of $\En$ from Eq.~\eqref{En_cc} calculated with Eq.~\eqref{Cd_approx} and Eq.~\eqref{C1_approx} for $N=16$ and the range $r = 3$ (red), 4 (blue), 5 (black). From these examples, we see that the proposed approach via the asymptotic spin-inversion expansion works sufficiently well for the correlator $\En$ as it captures the short time behaviour as well as the approximate time of crossing the Bell limit for $r\geqslant 4$. For $r=3$ the theory quantitatively shows that the limit is not surpassed.

\subsubsection{Gaussian approximation}\label{app:gaussian}
To gain further insight into the expansion method, we approximate the powers of cosines by Gaussian functions, sines with the first order in $\tau$, and $K-3\approx K$ in the second term of $|C_+|^2$, which is valid for large $N$. In this way, we obtain
\begin{align}
    |C_-|^2 &\approx P_r^2 \tau^N e^{-(K-N/2) \tau^2}, \\
    |C_+|^2 &\approx e^{-K \tau^2}( 1 + R_r^2 \tau^6).
\end{align}
Furthermore, in the region of $\tau$ where the Bell inequality is violated, and around the maximum of $\En$ we can neglect the first term in $|C_+|^2$ which yields our intermediate-time approximation 
\begin{align} \label{en.approx}
    \En  &\approx P_r^2 R_r^2 e^{-\beta_N \tau^2 } \tau^{N+6},
\end{align}
where we introduced the exponent $\beta_N = 2K - N/2$. 

Now, we estimate the time of crossing the Bell limit, and, as a result, we show that the approximate expression in Eq.~\eqref{en.approx} describes the observed numerical behaviour, i.e., that for $r=3$ the Bell threshold is not surpassed and for $r \geqslant 4$ we observe the violation of the inequality. However, our approach, being asymptotic in nature, does not capture accurately the critical time $\tau_c$, where the Bell limit is surpassed, due to its systematic underestimation which manifests in a small offset. 
In Fig.~\ref{fig:fig5}, this offset is not visible but it can be larger for larger values of $r$.

To determine $\tau_c$, we expand $\ln(2^N\En)$ around maximum, which is reached at the time $\tau_\mathrm{max} = \sqrt{(N+6)/(2\beta_N)}$, up to quadratic terms in $\tau-\tau_\mathrm{max}$. The Bell limit $\En=1/2^N$ is reached at
\begin{equation}\label{tcrit.approx}
    \tau_c \approx \tau_\mathrm{max} - \sqrt{ 
    \frac{\tilde{\mathcal{E}}_\mathrm{max}}{2\beta_N}
    },
\end{equation}
which is meaningful, i.e., $\tau_c \in \mathbb{R}$, only when $\tilde{\mathcal{E}}_\mathrm{max} \geqslant 0$, and
where the maximum of the scaled logarithm is
\begin{equation} \label{Emax.approx}
\frac{\tilde{\mathcal{E}}_\mathrm{max}}{N} = 
    \frac{2\ln(P_rR_r)}{N} 
    - \bigg(1+\frac{6}{N}\bigg) \ln\frac{1}{\tau_\mathrm{max}}
    +\ln2 - \frac{1+\frac{6}{N}}{2}.
\end{equation}
Due to the linear scaling of $R_r$ with $N$, the first term on the right-hand side extracts the exponent of $P_r$ in the first approximation. The second term, which decreases $\tilde{\mathcal{E}}_\mathrm{max}$, comes from the factor $\tau^N$ in the expansion of  $C_-$ originating from the requirement that all the $N$ spins have to be inverted during the dynamics. In Fig.~\ref{fig:fig5}(b), we present the function $\tilde{\mathcal{E}}_\mathrm{max}$ as a function of the range $r$ for various $N$'s.
The time $\tau_\mathrm{crit}$ from Eq.~\eqref{tcrit.approx}, when the Bell threshold is reached by $\En$, exists if $\tilde{\mathcal{E}}_\mathrm{max} \geqslant 0$. We find that for $r \leqslant 3$, $\tilde{\mathcal{E}}_\mathrm{max}$ is negative and for $r\geqslant 4$ it is positive resulting in a physically meaningful $\tau_c$.

Physically, $\En(\tau)$ quantifies how fast the initial state, given by $|C_+(\tau)|^2$, is depleted, and in order to surpass the Bell inequality, the state with all the spins inverted, given by $|C_-(\tau)|^2$, has to be populated fast enough. The population in $|C_-|^2$ is exponentially enhanced as $r$ increases due to a larger number of diagrams with spin-pairs inverted by interaction. On the other hand, $\tau_c$ decreases with increasing $r$ causing significant reduction $\propto \tau^N$ of $|C_-(\tau)|^2$. The two processes compete, and in order to surpass the Bell limit, $P_r$ has to be large enough. We note that $R_r$ also plays a role in the above considerations, since the first term in the expansion of $|C_+|^2$, which leads to Gaussian decay, is insufficient to describe the violation of the Bell limit.

\subsection{Measuring many-body Bell correlator with the Multiple Quantum Coherences}\label{sec:MQC}

The multiple quantum coherences (MQC) technique provides extensive information about the structure of a many-body state. It allows for the relation of the quantum correlators with other physical quantities like the out-of-time-order correlations~\cite{garttner2017measuring,PhysRevLett.120.040402}. The MQC is defined upon picking some observable $\hat A$ and expressing a density operator in terms of the eigenstates and eigenvalues of $\hat A$ as follows:
\begin{equation}
\hat\varrho=\sum_m\sum_{\lambda_i-\lambda_j=m}\varrho_{ij}\ketbra{\psi_i}{\psi_j}\equiv\sum_m\hat\varrho_m
\end{equation}
where $\hat{A} \ket{\psi_{i/j}}=\lambda_{i/j}\ket{\psi_{i/j}}$. 
The MQC is defined as a norm of the fixed-$m$ part of the density matrix, namely $I_m(\hat\varrho)=\tr{\hat\varrho^\dagger_m\hat\varrho_m}$.
In the multi-qubit case considered in the current work, we consider 
\begin{equation}
\hat A=\frac12\sum_{k=1}^N\hat\sigma_x^{(k)}    
\end{equation}
and $m=N/2$ to obtain $I_N(\hat\varrho)=\En$.
The $I_N(\hat\varrho)$ is directly accessible in the laboratory, as shown in Ref.~\cite{garttner2017measuring} for $N=6$ qubits. 

For a more detailed discussion of the physical significance of the MQCs, their measurements, and the relation to the out-of-time-order correlations, see Ref.~\cite{PhysRevLett.120.040402} and the references therein.

\putbib
\end{bibunit}

\end{document}